\documentclass[12pt]{article}
\usepackage{amssym}

\renewcommand{\theequation}{\arabic{section}.\arabic{equation}}
\newcommand{\pb}{\mbox{\boldmath $p$}}
\newcommand{\xxb}{\mbox{\boldmath $X$}}
\newcommand{\ppb}{\mbox{\boldmath $P$}}
\newcommand{\halpha}{\mbox{\boldmath $\hat{\alpha}$}}
\newcommand{\hbeta}{\hat{\beta}}
\newcommand{\sigmab}{\mbox{\boldmath $\sigma$}}
\newcommand{\llb}{\mbox{\boldmath $L$}}
\newcommand{\ssb}{\mbox{\boldmath $S$}}
\newcommand{\jjb}{\mbox{\boldmath $J$}}
\newcommand{\tphi}{\tilde{\phi}}
\newcommand{\trr}{\tilde{R}}
\def\case#1#2{{\textstyle{#1\over #2}}}
\newcommand{\tnn}{\tilde{N}}
\newcommand{\ta}{\tilde{a}}
\newcommand{\tb}{\tilde{b}}
\newcommand{\tn}{\tilde{n}}

\title{
Dirac oscillator with nonzero minimal uncertainty in position}
\author{C Quesne$^{1,3}$  and  V M Tkachuk$^2$\\
$^1$ {\small Physique Nucl\'eaire Th\'eorique et Physique
Math\'ematique,  Universit\'e Libre de Bruxelles,} \\ 
{\small Campus de la Plaine CP229, Boulevard~du Triomphe, B-1050 Brussels,
Belgium}\\ 
$^2$ {\small Ivan Franko Lviv National University, Chair of Theoretical
Physics,}\\
{\small 12, Drahomanov Street, Lviv UA-79005, Ukraine}\\
{\small E-mail: cquesne@ulb.ac.be and tkachuk@ktf.franko.lviv.ua}}
\date{ }
\begin{document}
\baselineskip=20pt plus 1pt minus 1pt
%%%%%%%%%%%%%%%%%%%%%%%%%%%%%%%%%%%%%%%%%%%%%%%%%%%%%%%%%%
\maketitle
\begin{abstract}
In the context of some deformed canonical commutation relations leading to isotropic nonzero minimal
uncertainties in the position coordinates, a Dirac equation is exactly solved for the first time, namely that
corresponding to the Dirac oscillator. Supersymmetric quantum mechanical and shape-invariance
methods are used to derive both the energy spectrum and wavefunctions in the momentum
representation. As for the conventional Dirac oscillator, there are neither negative-energy states for
$E=-1$, nor symmetry between the $l = j - \frac{1}{2}$ and $l = j + \frac{1}{2}$ cases, both features
being connected with supersymmetry or, equivalently, the $\omega \to - \omega$ transformation. In
contrast with the conventional case, however, the energy spectrum does not present any degeneracy
pattern apart from that associated with the rotational symmetry. More unexpectedly, deformation leads
to a difference in behaviour between the $l = j - \frac{1}{2}$ states corresponding to small, intermediate
and very large $j$ values in the sense that only for the first ones supersymmetry remains unbroken, while
for the second ones no bound state does exist. 
\end{abstract}

\vspace{0.5cm}



\vfill\noindent{\small $^3$ Research Director, National Fund for Scientific Research (FNRS), Belgium}.
%
%=========================================================================
%
\newpage
\section{Introduction}

Many years ago, It\^o et al~\cite{ito} considered a Dirac equation in which the momentum $\ppb$ is
replaced by $\ppb - {\rm i} m \omega \hbeta \xxb$ with $\xxb$ being the position vector, $m$ the
mass of the particle and $\omega$ the oscillator frequency. It was later shown by Cook~\cite{cook} to
present unusual accidental degeneracies in its spectrum, which were discussed from a supersymmetric
viewpoint by Ui and Takeda~\cite{ui} and by Balantekin~\cite{balantekin}.\par
%
%--------------------------------------------------------------------------------------
%
The interest in the problem was revived by Moshinsky and Szczepaniak~\cite{moshinsky89}, who gave
it the name Dirac oscillator (DO) because in the nonrelativistic limit it becomes a harmonic oscillator
with a very strong spin-orbit coupling term. Since then, the DO has aroused a lot of interest both
because it provides one of the few examples of Dirac equation exact solvability and because of its
numerous physical applications.\par
%
%----------------------------------------------------------------------------------
%
As a relativistic quantum mechanical problem, the DO has been studied from many viewpoints, including
covariance properties~\cite{moreno}, complete energy spectrum and corresponding
wavefunctions~\cite{benitez}, symmetry Lie algebra~\cite{cq90}, shift operators~\cite{delange},
hidden supersymmetry~\cite{benitez, beckers, cq91}, conformal invariance properties~\cite{martinez},
as well as completeness of wavefunctions~\cite{szmyt}.\par
%
%---------------------------------------------------------------------------------
%
Relativistic many-body problems with DO interactions have been extensively studied with special
emphasis on the mass spectra of mesons (quark-antiquark systems) and baryons (three-quark systems)
(see, e.g., \cite{moshinsky93a, moshinsky93b} and references quoted therein). The dynamics of
wavepackets in a DO has been determined~\cite{toyama, rozmej} and a relation with the
Jaynes-Cummings model established~\cite{rozmej}. The $2+1$ space time has also been
shown~\cite{villalba} to be an interesting framework for discussing the DO in connection with new
phenomena (such as the quantum Hall effect and fractional statistics) in condensed matter physics.
Thermodynamic properties~\cite{pacheco} of the DO in $1+1$ space time~\cite{dominguez} have been
mentioned to be relevant to studies on quark-gluon plasma models.\par
%
%-------------------------------------------------------------------------------
%
Various extensions of the DO have been considered in the literature. Let us mention some alternative
approach based on a scalar coupling~\cite{dixit}, as well as some generalizations to arbitrary
spin~\cite{moshinsky96} or to quasi-exactly solvable systems~\cite{ho}.\par
%
%--------------------------------------------------------------------------------
%
In the present paper, we plan to solve the DO problem in an entirely different context, namely that of
theories assuming a modified Heisenberg uncertainty relation leading to a nonzero minimal uncertainty in
position. Such theories have been stimulated by several independent lines of investigation in string
theory and quantum gravity suggesting the existence of a finite lower bound to the possible resolution
of length $\Delta X_0$ (see, e.g., \cite{gross, maggiore93a, witten}). In particular, we shall be interested in
a realization of this modified uncertainty relation through small quadratic corrections to the canonical
commutation relations~\cite{kempf94a, kempf97}. Another interesting aspect of these corrections is
that they can also provide an effective description of non-pointlike particles, such as quasiparticles and
various collective excitations in solids, or composite particles, such as nucleons and
nuclei~\cite{kempf97}.\par
%
%--------------------------------------------------------------------------------------
% 
Only a few nonrelativistic quantum mechanical problems have been considered in such a context so far. 
An exact solution to the one-dimensional harmonic oscillator problem has been provided by solving the
corresponding deformed Schr\"odinger equation in momentum representation~\cite{kempf95}. This
approach has been  extended to $D$ dimensions~\cite{chang}. Some perturbative~\cite{brau} or
partial~\cite{akhoury} results have also been obtained for the hydrogen atom.\par
%
%---------------------------------------------------------------------------------
%
The oscillator results, both in one and $D$ dimensions, have been rederived~\cite{cq03a, cq03b} by
using powerful combined techniques of supersymmetric quantum mechanics (SUSYQM)~\cite{cooper,
junker} and shape invariance (SI) under parameter translation~\cite{gendenshtein, dabrowska}. Such an
approach is known~\cite{junker, carinena} to be a reformulation of the factorization method dating back
to Schr\"odinger~\cite{schrodinger} and Infeld and Hull~\cite{infeld}. The combined formalism of
SUSYQM and SI (this time under parameter scaling~\cite{spiridonov, khare}) has proved especially
useful~\cite{cq03a, cq03b} when one assumes nonzero minimal uncertainties in both position and
momentum~\cite{kempf94b}, in which case neither position nor momentum representation can be
used.\par
%
%-----------------------------------------------------------------------------
%
Here we will avail ourselves of similar kinds of techniques to provide an exact solution to the DO problem
with nonzero minimal uncertainty in position. In section~2, this problem is reviewed in momentum
representation. In section~3, the equations for the large and small component radial momentum
wavefunctions are derived and the corresponding SUSY partners obtained. The existence of physically
acceptable wavefunctions corresponding to a vanishing energy for one of them is discussed in section~4. The
DO energy spectrum and wavefunctions are then determined in sections~5 and 6, respectively. The final
results are collected in section~7, while section~8 contains the conclusion.\par
%
%==================================================
%
\section{Dirac oscillator in momentum representation}

In a system of units wherein $\hbar = c = m =1$, the (stationary) DO equation is given
by~\cite{moshinsky89}
\begin{equation}
  H \psi = E \psi \qquad H = \halpha \cdot (\ppb - {\rm i} \omega \xxb \hbeta) + \hbeta
  \label{eq:DO-eq}
\end{equation}
where a hat denotes a $4 \times 4$ matrix,
\begin{equation}
  \halpha = \left(\begin{array}{cc}
      0 & \sigmab \\
      \sigmab & 0 
      \end{array}\right) \qquad
  \hbeta = \left(\begin{array}{cc}
      I & 0 \\
      0 & -I
      \end{array}\right) 
\end{equation}
and $\sigma_i$, $i=1$, 2, 3, are the Pauli spin matrices. Here we assume that the position and
momentum components $X_i$, $P_i$, $i=1$ 2, 3, satisfy modified commutation relations of the
type~\cite{kempf97, kempf95, chang}
\begin{eqnarray}
  [X_i, P_j] & = & {\rm i} \left[\delta_{i,j} (1 + \beta P^2) + \beta' P_i P_j \right]\nonumber \\{}
  [P_i, P_j] & = & 0  \label{eq:def-com} \\{}
  [X_i, X_j] & = & - {\rm i} \left[2\beta - \beta' + (2\beta + \beta') \beta P^2\right]
        \epsilon_{ijk} L_k\nonumber 
\end{eqnarray}
where there is a summation over dummy indices,
\begin{equation}
  L_i = \frac{1}{1 + \beta P^2} \epsilon_{ijk} X_j P_k \qquad i=1, 2, 3
\end{equation}
are the components of the orbital angular momentum, satisfying the usual commutation relations
\begin{equation}
  [L_i, X_j] = {\rm i} \epsilon_{ijk} X_k \qquad [L_i, P_j] = {\rm i} \epsilon_{ijk} P_k 
\end{equation}
and $\beta$, $\beta'$ are two nonnegative, very small deforming parameters. In the DO problem, we shall
assume that the two lengths $\sqrt{\beta}$ and $\sqrt{\beta'}$, which are the square roots of the
deforming parameters, are very small as compared with the oscillator characteristic length
$1/\sqrt{\omega}$ or, equivalently, that
$\beta \omega \ll 1$ and $\beta' \omega \ll 1$.\par
%
%--------------------------------------------------------------------------------------
%
The commutation relations (\ref{eq:def-com}) generalize to three dimensions (see, e.g., \cite{kempf97} for a
derivation in $d$ dimensions) the modified commutation relation between position and momentum in one
dimension
\begin{equation}
  [X, P] = {\rm i} (1 + \beta P^2).
\end{equation}
The latter has been proved~\cite{kempf95} to lead (in those states for which $\langle P \rangle = 0$) to the
uncertainty relation
\begin{equation}
  \Delta X \ge \frac{1}{2} \left(\frac{1}{\Delta P} + \beta\, \Delta P \right)  \label{eq:UR}
\end{equation}
appearing in perturbative string theory and in line with the proposed UV-IR mixing~\cite{gross, maggiore93a,
witten}. Equation (\ref{eq:UR}) implies a nonzero lower bound for $\Delta X$: $\Delta X \ge \Delta X_0 =
\sqrt{\beta}$.\par
%
%------------------------------------------------------------------------------------------------------
%
{}For the commutation relations (\ref{eq:def-com}), the same mechanism as in one dimension yields from
the corresponding uncertainty relation $\Delta X_i \Delta P_i \ge \frac{1}{2} |\langle[X_i, P_i]\rangle|$ an
isotropic nonzero minimal uncertainty $\Delta X_0 = \Delta X_{0i}$, $i=1$, 2, 3, equal to $\Delta X_0 =
\sqrt{3\beta + \beta'}$ when one restricts oneself to those states for which $\langle P_i\rangle = 0$,
$i=1$, 2, 3, and $\Delta P_i$ is independent of $i$~\cite{kempf97}.\par
%
%-----------------------------------------------------------------------------------------------------
%  
Since, from (\ref{eq:def-com}), it follows that the momentum components remain simultaneously
diagonalizable, we can work in the momentum representation, wherein $P_i$, $X_i$ and $L_i$ are realized
as~\cite{chang}
\begin{eqnarray}
  P_i & = & p_i  \nonumber \\
  X_i & = & {\rm i} \left[(1 + \beta p^2) \frac{\partial}{\partial p_i} + \beta' p_i p_j
       \frac{\partial}{\partial p_j} + \gamma p_i\right]  \label{eq:mom-rep} \\
  L_i & = & {\rm i} \epsilon_{ijk} \frac{\partial}{\partial p_j} p_k = - {\rm i} \epsilon_{ijk} p_j
       \frac{\partial}{\partial p_k}. \nonumber  
\end{eqnarray}
Here $\gamma$ is an arbitrary constant, which does not appear in the commutation relations
(\ref{eq:def-com}) and only affects the weight function in the scalar product in momentum space
\begin{equation}
  \langle \zeta' | \zeta\rangle = \int \frac{d^3 \pb}{[f(p)]^{1-\alpha}} \zeta^{\prime*}(\pb)
  \zeta(\pb)  \label{eq:sp}
\end{equation}
where
\begin{equation}
  f(p) \equiv 1 + \beta_0 p^2 \qquad \beta_0 \equiv \beta + \beta' \qquad \alpha \equiv
  \frac{\gamma - \beta'}{\beta_0}.  \label{eq:f}
\end{equation}
\par
%
%----------------------------------------------------------------------------------
%
On separating the wavefunction $\psi = \left( \begin{array}{c} \psi_1 \\ \psi_2\end{array}\right)$ in
(\ref{eq:DO-eq}) into large $\psi_1$ and small $\psi_2$ components, the DO equation can be written
as two coupled equations
\begin{eqnarray}
  B^+ \psi_2 & =& (E-1) \psi_1  \label{eq:DO-eq1} \\
  B^- \psi_1 & =& (E+1) \psi_2  \label{eq:DO-eq2}
\end{eqnarray}
where
\begin{equation}
  B^{\pm} = \sigmab \cdot \ppb \pm {\rm i} \omega \sigmab \cdot \xxb.  \label{eq:B-def}
\end{equation}
Applying $B^+$ (resp.\ $B^-$) to (\ref{eq:DO-eq2}) (resp.\ (\ref{eq:DO-eq1})) and using
(\ref{eq:DO-eq1}) (resp.\ (\ref{eq:DO-eq2}), we get the following factorized equation for the large
component $\psi_1$ (resp.\ small component $\psi_2$)
\begin{eqnarray}
  B^+ B^- \psi_1 & = & (E^2-1) \psi_1 \\
  B^- B^+ \psi_2 & = & (E^2-1) \psi_2.
\end{eqnarray}
\par
%
%----------------------------------------------------------------------------------------
%
As shown in appendix 1, the operators $B^+$ and $B^-$ can be written in the momentum
representation (\ref{eq:mom-rep}) as
\begin{eqnarray}
  B^+ & = & \omega \left\{\frac{p}{\omega} - \left[(1 + \beta p^2) \left(\frac{\partial}{\partial p} +
       \frac{\sigmab\cdot\llb+2}{p}\right) + \beta' p^2 \frac{\partial}{\partial p} + \gamma p\right]
       \right\} \sigma_p  \label{eq:B+} \\ 
  B^- & = & \omega \sigma_p \left\{\frac{p}{\omega} + \left[(1 + \beta p^2)
       \left(\frac{\partial}{\partial p} - \frac{\sigmab\cdot\llb}{p}\right) + \beta' p^2
       \frac{\partial}{\partial p} + \gamma p\right]\right\}  \label{eq:B-} 
\end{eqnarray}
where
\begin{equation}
  \sigma_p \equiv \frac{\sigmab \cdot \pb}{p}  \label{eq:sigmap}
\end{equation}
is such that
\begin{equation}
  \sigma_p^2 = I.  \label{eq:sigmap2}
\end{equation}
\par
%
%---------------------------------------------------------------------------------------
%
Some further simplifications are achieved by transforming the large and small components according to
\begin{equation}
  \psi_i = \frac{1}{p} f^{-\alpha/2} \phi_i \qquad i=1, 2  \label{eq:psi-phi}
\end{equation}
where $f$ and $\alpha$ are defined in (\ref{eq:f}). As a consequence, equations (\ref{eq:DO-eq1}) and
(\ref{eq:DO-eq2}) become
\begin{eqnarray}
  {\cal B}^+ \phi_2 & = & (E-1) \phi_1 \label{eq:DO-eq1bis}\\
  {\cal B}^- \phi_1 & = & (E+1) \phi_2  \label{eq:DO-eq2bis} 
\end{eqnarray}
where ${\cal B}^{\pm}$ denote the transformed operators $B^{\pm}$, which can be written as
\begin{eqnarray}
  {\cal B}^+ & = & \omega b^+ \sigma_p = \omega \sigma_p \tilde{b}^+ \label{eq:tB+} \\ 
  {\cal B}^- & = & \omega \sigma_p b^- = \omega \tilde{b}^- \sigma_p. \label{eq:tB-} 
\end{eqnarray}
Here the operators $b^{\pm}$ are directly obtained from (\ref{eq:f}), (\ref{eq:B+}), (\ref{eq:B-}) and
(\ref{eq:psi-phi}) as
\begin{equation}
  b^{\pm} = \mp f \frac{\partial}{\partial p} + \left(\frac{1}{\omega} - \beta (\sigmab \cdot \llb +
  1)\right) p - \frac{\sigmab \cdot \llb + 1}{p}  \label{eq:b+-}
\end{equation}
while the alternative forms
\begin{equation}
  \tilde{b}^{\pm} = \mp f \frac{\partial}{\partial p} + \left(\frac{1}{\omega} + \beta (\sigmab \cdot 
  \llb + 1)\right) p + \frac{\sigmab \cdot \llb + 1}{p}
\end{equation}
result from (\ref{eq:b+-}) and the relation
\begin{equation}
  \{\sigma_p, \sigmab \cdot \llb + 1\} = 0  \label{eq:anticom}
\end{equation}
proved in appendix 1. From (\ref{eq:sigmap2}), (\ref{eq:tB+}) and (\ref{eq:tB-}), it follows that
equations (\ref{eq:DO-eq1bis}) and (\ref{eq:DO-eq2bis}) can be rewritten as
\begin{eqnarray}
  \omega b^+ \tphi_2 & = & (E-1) \phi_1 \label{eq:DO-eq1ter} \\
  \omega b^- \phi_1 & = & (E+1) \tphi_2  \label{eq:DO-eq2ter}
\end{eqnarray} 
where
\begin{equation}
  \tphi_2 = \sigma_p \phi_2.  \label{eq:tphi-phi}
\end{equation}
\par
%
%===================================================
%
\section{Large and small component radial momentum wavefunctions and SUSY partners}
\setcounter{equation}{0}

In section 2, we have demonstrated that the solution to the DO equation (\ref{eq:DO-eq}) with modified
commutation relations (\ref{eq:def-com}) amounts to that of the system of coupled equations
(\ref{eq:DO-eq1ter}) and (\ref{eq:DO-eq2ter}). Since the operators $b^{\pm}$ appearing on the
left-hand side of such equations have the property of commuting with the total angular momentum
$\jjb = \llb + \ssb$, where $\ssb = \frac{1}{2} \sigmab$, as well as with $\llb^2$ and $\ssb^2$, we
may look for solutions $\phi_1$ and $\tphi_2$ that are simultaneous eigenfunctions of $\llb^2$,
$\ssb^2$, $\jjb^2$ and $J_3$, corresponding to the eigenvalues $l(l+1)$, $\frac{3}{4}$, $j(j+1)$
and $m$, respectively. Instead of $l$, we may use $s$, defined by $l = j-s$ and taking the values
$\pm \frac{1}{2}$. Then $\phi_1$ and $\tphi_2$ can be expressed in momentum spherical
coordinates $p$, $\theta$, $\varphi$ and spin variable $\xi$ as
\begin{eqnarray}
  \phi_1 & = & \phi_{1;s,j,m}(p, \theta, \varphi, \xi) = R_{1;s,j}(p) {\cal Y}_{s,j,m}(\theta, \varphi,
       \xi) \label{eq:phi1}\\
  \tphi_2 & = & \tphi_{2;s,j,m}(p, \theta, \varphi, \xi) = \trr_{2;s,j}(p) {\cal Y}_{s,j,m}(\theta, \varphi,
       \xi)  \label{eq:tphi2}
\end{eqnarray}  
where
\begin{equation}
  {\cal Y}_{s,j,m}(\theta, \varphi, \xi) = \sum_{\mu, \sigma} \langle j-s \;\mu, \case{1}{2} \;
  \sigma \mid j \;m\rangle Y_{j-s,\mu}(\theta, \varphi) \chi_{\sigma}(\xi)  \label{eq:sp-harmo}
\end{equation}
is a spin spherical harmonic~\cite{edmonds} and $R_{1;s,j}(p)$, $\trr_{2;s,j}(p)$ (or, in short,
$R_1(p)$, $\trr_2(p)$) are radial wavefunctions. Note that in (\ref{eq:sp-harmo}), $\chi_{\sigma}(\xi)$,
where $\sigma = \pm \frac{1}{2}$, denotes a spinor. \par
%
%------------------------------------------------------------------------------------
% 
We note that $\phi_1$ and $\tphi_2$ have the same angular-spin wavefunction. The same is not true
for $\phi_1$ and $\phi_2$ since, from (\ref{eq:sigmap2}) and (\ref{eq:tphi-phi}), the latter is given by
\begin{equation}
  \phi_2 = \sigma_p \tphi_2 = \trr_{2;s,j}(p) \sigma_p {\cal Y}_{s,j,m}(\theta, \varphi, \xi) =
  - \trr_{2;s,j}(p) {\cal Y}_{-s,j,m}(\theta, \varphi, \xi)
\end{equation}
where, in the last step, we used a well-known property of spin spherical harmonics~\cite{rose}. Hence
we may write
\begin{equation}
  \phi_2 = \phi_{2;-s,j,m}(p, \theta, \varphi, \xi) = R_{2;-s,j}(p) {\cal Y}_{-s,j,m}(\theta, \varphi,
       \xi)  \label{eq:phi2}
\end{equation}
where 
\begin{equation}
  R_{2;-s,j}(p) = - \trr_{2;s,j}(p).  \label{eq:R-Rtilde}
\end{equation}
In other words, $\phi_1$ and $\tphi_2$ are characterized by the same $l$ value ($l=j-s$), whereas
$\phi_2$ corresponds to $l'=j+s$.\par
%
%--------------------------------------------------------------------------------------
%
Inserting (\ref{eq:phi1}) and (\ref{eq:tphi2}) in (\ref{eq:DO-eq1ter}) and (\ref{eq:DO-eq2ter}) and
taking the property 
\begin{equation}
  (\sigmab \cdot \llb + 1) {\cal Y}_{s,j,m} = (\jjb^2 - \llb^2 - \ssb^2 + 1) {\cal Y}_{s,j,m} = s(2j+1)
  {\cal Y}_{s,j,m} 
\end{equation}
into account, we obtain a system of coupled radial equations
\begin{eqnarray}
  \omega b_p^+ \trr_2 = (E-1) R_1  \label{eq:DO-eq1quater} \\
  \omega b_p^- R_1 = (E+1) \trr_2.  \label{eq:DO-eq2quater}
\end{eqnarray}
Here $b_p^{\pm}$ denote the radial parts of the operators $b^{\pm}$
\begin{equation}
  b_p^{\pm} = b_p^{\pm}(g,k) = \mp f \frac{d}{dp} + gp - \frac{k}{p}  \label{eq:bp}
\end{equation}
where we have introduced the notations
\begin{equation}
  g = \frac{1}{\omega} - \beta s (2j+1) \qquad k = s (2j+1).  \label{eq:g-k}
\end{equation}
\par
%
%---------------------------------------------------------------------------------------
%
According to (\ref{eq:psi-phi}), (\ref{eq:phi1}) and (\ref{eq:tphi2}), the scalar product (\ref{eq:sp}) in
momentum space gives rise to a scalar product
\begin{equation}
  \langle R' | R\rangle = \int_0^{\infty} \frac{dp}{f(p)} R^{\prime*}(p) R(p)  \label{eq:sp-bis}
\end{equation}
in radial momentum space. It is easily checked that with respect to the latter the operators $b_p^+$
and $b_p^-$, defined in (\ref{eq:bp}), are Hermitian conjugates of one another.\par
%
%------------------------------------------------------------------------------------
%
{}Finally, from (\ref{eq:DO-eq1quater}) and (\ref{eq:DO-eq2quater}), we get
\begin{eqnarray}
  b_p^+ b_p^- R_1 & = & e R_1  \label{eq:SUSY1}  \\
  b_p^- b_p^+ \trr_2 & = & e \trr_2  \label{eq:SUSY2}
\end{eqnarray}
where
\begin{equation}
  e \equiv \frac{E^2-1}{\omega^2}.
\end{equation}
We conclude that $R_1$ and $\trr_2$ may be considered as eigenfunctions of two SUSY partner
Hamiltonians~\cite{cooper, junker}. Before using this property to solve equations (\ref{eq:SUSY1}) and
(\ref{eq:SUSY2}) in full generality, it is worth reviewing the conditions that should be imposed on $R_1$
and $\trr_2$ to make them physically acceptable.\par
%
%--------------------------------------------------------------------------------------------
%
As in any bound-state problem for the Dirac equation, the complete relativistic wavefunction $\psi = \left(
\begin{array}{c} \psi_1 \\ \psi_2 \end{array}\right)$ should be normalizable, which in the present case
amounts to the condition
\begin{equation}
  \int_0^{\infty} \frac{dp}{f(p)} \left(\Bigl|R_1(p)\Bigr|^2 + \left|\trr_2(p)\right|^2\right) = 1
  \label{eq:normalization}
\end{equation}
implying separate convergence of (\ref{eq:sp-bis}) for $R(p) = R'(p) = R_1(p)$ and $R(p) = R'(p) =
\trr_2(p)$.\par
%
%-----------------------------------------------------------------------------------------------
%
In deformed quantum mechanics governed by the generalized commutation relations
(\ref{eq:def-com}), it has been shown, however, that normalizability may not be a sufficient
condition for a wavefunction to be physically acceptable~\cite{kempf95}: it should indeed be in the
domain of $\ppb$, which physically means that it should have a finite uncertainty in momentum. This
leads to imposing 
\begin{equation}
  \langle p^2\rangle = \int_0^{\infty} \frac{dp}{f(p)} p^2 \left(\Bigl|R_1(p)\Bigr|^2 +
  \left|\trr_2(p)\right|^2\right) < \infty  \label{eq:p2}
\end{equation}
and therefore
\begin{equation}
  \int_0^{\infty} \frac{dp}{f(p)} p^2 \Bigl|R_1(p)\Bigr|^2 < \infty \qquad \int_0^{\infty} \frac{dp}{f(p)}
  p^2 \left|\trr_2(p)\right|^2 < \infty   \label{eq:p2-bis}
\end{equation}
for the large and small components, respectively.\par
%
%----------------------------------------------------------------------------------------
%
In the next section, we will proceed to determine whether such conditions can be fulfilled for the minimal
$e$ value, i.e., $e=0$ or $E^2=1$.\par
%
%==============================================
% 
\section{\boldmath Ground state with $E^2=1$ or $e=0$ for given $s$ and $j$ values}
\setcounter{equation}{0}

{}From the Hermiticity properties of $b_p^+$ and $b_p^-$ with respect to the scalar product
(\ref{eq:sp-bis}), it results that
\begin{eqnarray}
  \langle R_1 | b_p^+ b_p^- | R_1\rangle & = & \langle b_p^- R_1 | b_p^- R_1\rangle \\
  \langle \trr_2 | b_p^- b_p^+ | \trr_2\rangle & = & \langle b_p^+ \trr_2 | b_p^+ \trr_2\rangle.
\end{eqnarray}
Hence the solutions of equations (\ref{eq:SUSY1}) and (\ref{eq:SUSY2}) corresponding to $e=0$ are
necessarily solutions of the first-order equations
\begin{eqnarray}
  b_p^- R_1^{(0)} & = & 0  \\
  b_p^+ \trr_2^{(0)} & = & 0
\end{eqnarray}
respectively. Provided they are physically acceptable, these are given by
\begin{eqnarray}
  R_1^{(0)} & = & N_1^{(0)} p^k f^{-(g + \beta_0 k)/(2\beta_0)}  \label{eq:gs-wf} \\
  \trr_2^{(0)} & = & \tnn_2^{(0)} p^{-k} f^{(g + \beta_0 k)/(2\beta_0)}
\end{eqnarray}
where $N_1^{(0)}$ and $\tnn_2^{(0)}$ are some normalization coefficients.\par
%
%---------------------------------------------------------------------------
%
Let us first consider the case of $R_1^{(0)}$ and examine the convergence of the corresponding
normalization integral (\ref{eq:sp-bis}). Since for $p \to 0$, $R_1^{(0)}$ behaves as
$p^k$, the boundary condition $R_1^{(0)}(0) = 0$ imposes the restriction $k>0$. From (\ref{eq:g-k}), it
follows that this can only be fulfilled for $s=\frac{1}{2}$. Furthermore, for $p \to \infty$, $R_1^{(0)}$
behaves as $p^{-g/\beta_0}$, so that the convergence of (\ref{eq:sp-bis}) will be ensured provided,
in addition,
\begin{equation}
  \frac{g}{\beta_0} > - \frac{1}{2}  \label{eq:norm-cond1}
\end{equation}
which, for $s=\frac{1}{2}$, is equivalent to the condition
\begin{equation}
  2\beta \omega j - \beta' \omega < 2.  \label{eq:small-j}
\end{equation}
Hence $R_1^{(0)}$ is normalizable only for $s=\frac{1}{2}$ and small $j$ values
satisfying (\ref{eq:small-j}).\par
%
%-----------------------------------------------------------------------------
%
It is however immediately clear that the convergence of the first integral in (\ref{eq:p2-bis}) imposes the
more stringent condition
\begin{equation}
  \frac{g}{\beta_0} > \frac{1}{2}  \label{eq:norm-cond1-bis}
\end{equation}
or, explicitly,
\begin{equation}
  2\beta \omega (j+1) + \beta' \omega < 2.  \label{eq:small-j-bis}
\end{equation}
We therefore conclude that for $R_1^{(0)}$ we must require both $s = \frac{1}{2}$ and small $j$ 
values fulfilling (\ref{eq:small-j-bis}).\par
%
%-----------------------------------------------------------------
%
Considering next the case of $\trr_2^{(0)}$ normalizability, from the boundary condition
$\trr_2^{(0)}(0) = 0$ we find this time the restriction $k<0$, equivalent to $s = - \frac{1}{2}$.
Moreover, from the behaviour for $p \to \infty$, we get the additional condition $g/\beta_0 < 1/2$,
which for $s = - \frac{1}{2}$ amounts to the inequality $2\beta \omega j - \beta' \omega < -2$.
Since for very small values of $\beta \omega$ and $\beta' \omega$, such a condition cannot be
satisfied by any $j$ value, we conclude that the function $\trr_2^{(0)}$ is never normalizable.\par
%
%----------------------------------------------------------------------------
%  
{}For those functions $R_1^{(0)}$ that are physically acceptable, we must therefore set $\trr_2^{(0)}
= 0$ for the corresponding small component radial momentum wavefunction. In other words, we have
proved that the set of equations (\ref{eq:SUSY1}), (\ref{eq:SUSY2}) has well-behaved solutions
$\left(R_1^{(0)} \ne 0, \trr_2^{(0)} = 0\right)$, corresponding to $e=0$, i.e., $E = \pm 1$, for $s =
\frac{1}{2}$ and any $j$ value satisfying condition (\ref{eq:small-j-bis}).\par
%
%-----------------------------------------------------------------------------
% 
When going back, however, to the coupled radial equations (\ref{eq:DO-eq1quater}),
(\ref{eq:DO-eq2quater}), from which the eigenvalue problems (\ref{eq:SUSY1}) and (\ref{eq:SUSY2})
have been derived, we note that the solutions $\left(R_1^{(0)} \ne 0, \trr_2^{(0)} = 0\right)$ are
only compatible with $E=1$ to the exclusion of $E=-1$. So, in the latter case, although the set of
equations (\ref{eq:SUSY1}), (\ref{eq:SUSY2}) has a well-behaved solution, it is unphysical in the sense
that it is not a solution of (\ref{eq:DO-eq1quater}), (\ref{eq:DO-eq2quater}).\par
%
%-------------------------------------------------------------------------------
%
In conclusion, we have shown that as in the case of standard commutation relations~\cite{benitez},
physically acceptable wavefunctions of (\ref{eq:DO-eq}) with $E^2=1$ can only exist for positive
energy and $s=\frac{1}{2}$, i.e., $E=1$ and $l = j - \frac{1}{2}$. In the present case of modified
commutation relations (\ref{eq:def-com}), there however occurs an additional condition, not present
in the standard case, namely that the total angular momentum $j$ be restricted to small values
satisfying  equation (\ref{eq:small-j-bis}).\par
%
%-----------------------------------------------------------------------------
%
In physical terms, this means that there will be neither particles nor antiparticles with $E=1$ and $j$ values
violating condition (\ref{eq:small-j-bis}), while this remark is only true for antiparticles whenever $j$ satisfies
(\ref{eq:small-j-bis}).\par
%
%=================================================
% 
\section{Dirac oscillator energy spectrum}
\setcounter{equation}{0}

We will now proceed to the solution of equations (\ref{eq:SUSY1}), (\ref{eq:SUSY2}) (and ultimately to
that of equations (\ref{eq:DO-eq1quater}), (\ref{eq:DO-eq2quater})). In accordance with section 4
results, we have to distinguish between the three cases (i) $s=\frac{1}{2}$, small $j$, (ii)
$s=\frac{1}{2}$, large $j$, and (iii) $s=-\frac{1}{2}$.\par
%
%+++++++++++++++++++++++++++++++++++++++++++++++++++
% 
\subsection{\boldmath Case $\frac{1}{2}$ and small $j$}

It has been shown in section 4 that whenever $j$ satisfies condition (\ref{eq:small-j-bis}), equation
(\ref{eq:SUSY1}) has a physically acceptable solution $R_1^{(0)}(g,k;p)$ corresponding to the ground
state energy $e_0=0$.\par
%
%-----------------------------------------------------------------------------------
% 
The other eigenvalues corresponding to $s=\frac{1}{2}$ and the same $j$ value can be found by
considering  $h_0 = b_p^+ b_p^-$ as the first member of a SUSYQM hierarchy~\cite{cooper, junker}
\begin{equation}
  h_i = b_p^+(g_i,k_i) b_p^-(g_i,k_i) + \sum_{j=0}^i \epsilon_j \qquad i=0, 1, 2, \ldots 
\end{equation}
where $g_i$, $k_i$, $i=0$, 1, 2,~\ldots, are some parameters satisfying the conditions
\begin{eqnarray}
  k_i & > & 0 \qquad \frac{g_i}{\beta_0} > \frac{1}{2}  \label{eq:SIcond1} \\
  g_0 & =& g \qquad k_0 = k \qquad \epsilon_0 = 0  \label{eq:SIcond2}
\end{eqnarray} 
and by imposing a SI condition~\cite{gendenshtein, dabrowska}
\begin{equation}
  b_p^-(g_i,k_i) b_p^+(g_i,k_i) = b_p^+(g_{i+1},k_{i+1}) b_p^-(g_{i+1},k_{i+1}) + \epsilon_{i+1}
  \qquad i=0, 1, 2, \ldots.  \label{eq:SI-cond} 
\end{equation}
Equation (\ref{eq:SI-cond}) explicitly reads
\begin{eqnarray}
  k_{i+1} (k_{i+1}-1) & = & k_i (k_i+1)  \label{eq:SI1} \\
  g_{i+1} (g_{i+1} - \beta_0) & = & g_i (g_i + \beta_0)  \label{eq:SI2} \\
  \epsilon_{i+1} & = & g_{i+1} (2k_{i+1}+1) - g_i (2k_i-1) + \beta_0 (k_{i+1} + k_i).  \label{eq:SI3}
\end{eqnarray}
\par
%
%---------------------------------------------------------------------------------------
%
The solution of the first two equations (\ref{eq:SI1}) and (\ref{eq:SI2}), satisfying conditions
(\ref{eq:SIcond1}) and (\ref{eq:SIcond2}), is given by
\begin{equation}
  k_i = k+i \qquad g_i = g + \beta_0 i.  \label{eq:ki}
\end{equation}
For such values of $g_i$ and $k_i$, the hierarchy Hamiltonians $h_i$ have physically acceptable
solutions $R_1^{(0)}(g_i,k_i;p)$ corresponding to the energies $\sum_{j=0}^i \epsilon_j$.\par
%
%-----------------------------------------------------------------------------
%
Using (\ref{eq:SI3}), we get for the eigenvalues of $h_0$ the result
\begin{equation}
  e_n = e_n(g,k) = \sum_{j=0}^i \epsilon_j = 4n [g + \beta_0 (k+n)].  \label{eq:en}
\end{equation}
Since, in the present case, equation (\ref{eq:g-k}) yields $g = \frac{1}{\omega} - \beta (j+\frac{1}{2})$
and $k = j+\frac{1}{2}$, the corresponding DO eigenvalues are given by
\begin{equation}
  E_n^2 - 1 = \omega^2 e_n = 4\omega n \left[1 - \beta \omega \left(j + \frac{1}{2}\right) +
  \beta_0 \omega \left(n + j + \frac{1}{2}\right)\right]. 
\end{equation}
In section 6, it will be proved that one can associate to these eigenvalues some physically acceptable
solutions $\left(R_1^{(n)}, \trr_2^{(n)}\right)$ of the coupled equations (\ref{eq:DO-eq1quater}),
(\ref{eq:DO-eq2quater}) for $n=0$, 1, 2,~\ldots\ or $n=1$, 2,~\ldots, according to whether the energy
is positive or negative, respectively.\par
%
%+++++++++++++++++++++++++++++++++++++++++++++++++++++
% 
\subsection{\boldmath Case $s = \frac{1}{2}$ and large $j$}

Whenever
\begin{equation}
  \frac{g}{\beta_0} < \frac{1}{2}  \label{eq:norm-cond2}
\end{equation}
or, in explicit form,
\begin{equation}
  2\beta \omega (j+1) + \beta' \omega > 2  \label{eq:large-j}
\end{equation}
there is no physically acceptable wavefunction $R_1^{(0)}(g,k;p)$ corresponding to $e=0$ or $E^2=1$.
We will now proceed to prove that for $j$ values satisfying condition (\ref{eq:large-j}), there
exists a ground state of
\begin{equation}
  h_0 = b_p^+(g,k) b_p^-(g,k) = - \left(f \frac{d}{dp}\right)^2 + g (g-\beta_0) p^2 + 
  \frac{k(k-1)}{p^2} -2gk - g - \beta_0 k  \label{eq:h0}
\end{equation}
with $e>0$ or $E^2>1$.\par 
%
%-----------------------------------------------------------------------------
%
{}For this purpose, it will prove convenient to re-factorize $h_0$. On introducing
\begin{equation}
  g' = - g + \beta_0  \label{eq:g'}
\end{equation}
equation (\ref{eq:h0}) can indeed be rewritten as
\begin{equation}
  h_0 = b_p^+(g',k) b_p^-(g',k) + (2k+1) (- 2g + \beta_0)
\end{equation}
where $b_p^{\pm}(g',k)$ are defined as in equation (\ref{eq:bp}) but with $g$ replaced by $g'$.\par
%
%--------------------------------------------------------------------------
%
If there exists a physically acceptable function $R_1^{(0)}(g',k;p)$ satisfying the equation
\begin{equation}
  b_p^-(g',k) R_1^{(0)} = 0
\end{equation}
then it will be the ground state wavefunction of $h_0$, corresponding to the eigenvalue
\begin{equation}
  e_0 = (2k+1) (- 2g + \beta_0).  \label{eq:e0}
\end{equation}
That this is indeed true follows from section 4 results: $R_1^{(0)}$ is given by
\begin{equation}
  R_1^{(0)} = N_1^{(0)} p^k f^{-(g' + \beta_0 k)/(2\beta_0)}
\end{equation}
where $g'/\beta_0 > 1/2$ as a consequence of (\ref{eq:norm-cond2}) and (\ref{eq:g'}), while $k>0$
since $s=\frac{1}{2}$. Hence $R_1^{(0)}(g',k;p)$ is both normalizable and such that its contribution to
$\langle p^2 \rangle$ converges.\par
%
%-----------------------------------------------------------------------
%
The remaining eigenvalues of $h_0$ can be directly obtained from (\ref{eq:en}) by substituting there
$g'$ for $g$, taking (\ref{eq:g'}) into account and adding the extra contribution (\ref{eq:e0}):
\begin{equation}
  e_n = 4n [- g + \beta_0 (k+1+n)] + (2k+1) (- 2g + \beta_0).
\end{equation}
On inserting the appropriate values of $g$ and $k$, the resulting DO eigenvalues read
\begin{equation}
  E_n^2 - 1 = \omega^2 e_n = 4\omega (n+j+1) \left[-1 + \beta \omega \left(j + \frac{1}{2}\right)
  + \beta_0 \omega \left(n + \frac{1}{2}\right)\right]. \label{eq:E-large}
\end{equation}
\par
%
%-----------------------------------------------------------------------------
%
The existence of such eigenvalues is however restricted to that of physically acceptable solutions
$\left(R_1^{(n)}, \trr_2^{(n)}\right)$ of equations (\ref{eq:DO-eq1quater}) and (\ref{eq:DO-eq2quater}).
In section~6, we will show that $\trr_2^{(0)} \ne 0$ contrary to what happens for small $j$ values and
$E_0=1$, but that the convergence of its contribution to $\langle p^2\rangle$ (i.e., the second condition
in (\ref{eq:p2-bis})) imposes a further restriction on the allowed $g$ values. This will lead us to distinguish
between very large $j$ values for which $n=0$, 1, 2,~\ldots\ are allowed in (\ref{eq:E-large}), irrespective of
the energy sign, and intermediate $j$ values for which no bound state does exist.\par
%
%+++++++++++++++++++++++++++++++++++++++++++++++++
%
\subsection{\boldmath Case $s = -\frac{1}{2}$}

Here the negative value of $k = -(j+\frac{1}{2})$ prevents the existence of a normalizable
wavefunction $R_1^{(0)}(g,k;p)$ corresponding to $e=0$ or $E^2=1$. As in the previous case,
however, it can be shown by re-factorization that $h_0$ has a ground state with $e>0$ or $E^2>1$.\par
%
%-------------------------------------------------------------------------------
%
On introducing
\begin{equation}
  k' = 1 - k = j + \frac{3}{2}
\end{equation}
equation (\ref{eq:h0}) can indeed be transformed into
\begin{equation}
  h_0 = b_p^+(g,k') b_p^-(g,k') + 2(2g + \beta_0) (j+1)
\end{equation}
where $b_p^{\pm}(g,k')$ are defined as in equation (\ref{eq:bp}) but with $k$ replaced by $k'$.\par
%
%-------------------------------------------------------------------------------------
%
The ground state wavefunction $R_1^{(0)}(g,k';p)$ of $h_0$, corresponding to the eigenvalue
\begin{equation}
  e_0 = 2 (2g + \beta_0) (j+1)
\end{equation}
is then provided by the solution of the equation
\begin{equation}
  b_p^-(g,k') R_1^{(0)} = 0
\end{equation}
which can be written as
\begin{equation}
  R_1^{(0)} = N_1^{(0)} p^{k'} f^{-(g + \beta_0 k')/(2\beta_0)}.
\end{equation}
Such a function is normalizable since now $k'>0$, while $g = \frac{1}{\omega} + \beta
(j+\frac{1}{2})$ satisfies condition (\ref{eq:norm-cond1}). In addition, for $\beta \omega \ll 1$
and $\beta' \omega \ll 1$, its contribution to $\langle p^2\rangle$ also converges since condition
(\ref{eq:norm-cond1-bis}) can now be written as
\begin{equation}
  2 + 2 \beta \omega j > \beta' \omega 
\end{equation}
which is fulfilled for any $j$.\par
%
%-------------------------------------------------------------------------------
% 
By proceeding in a similar way as in section 5.2, we find for the remaining eigenvalues of $h_0$ and the
corresponding DO eigenvalues
\begin{equation}
  e_n = 4n [g + \beta_0 (1-k+n)] + 2(2g + \beta_0) (j+1)
\end{equation}
and 
\begin{equation}
  E_n^2 - 1 = \omega^2 e_n = 4\omega (n+j+1) \left[1 + \beta \omega \left(j + \frac{1}{2}\right)
  + \beta_0 \omega \left(n + \frac{1}{2}\right)\right]
\end{equation}
respectively. In section~6, it will be shown that there exist physically acceptable solutions $\left(R_1^{(n)},
\trr_2^{(n)}\right)$ of equations (\ref{eq:DO-eq1quater}), (\ref{eq:DO-eq2quater}) for any $n=0$, 1,
2,~\ldots, irrespective of the energy sign, and that $\trr_2^{(0)} \ne 0$.\par
%
%================================================
%  
\section{Dirac oscillator radial momentum wavefunctions}
\setcounter{equation}{0}

In sections 4 and 5, the ground state wavefunction of $h_0 = b_p^+ b_p^-$ has been determined,
thus providing us with the large component radial momentum wavefunction for the lowest state with
given $s$ and $j$ values. In this section, we plan to calculate the remaining large and small component
radial momentum wavefunctions and to examine whether they are physically acceptable.\par
%
%++++++++++++++++++++++++++++++++++++++++++++++++++++++++
%
\subsection{\boldmath Case $s=\frac{1}{2}$ and small $j$}

The large component radial momentum wavefunctions for the excited states with given $s$ and $j$
values correspond to the excited state wavefunctions of $h_0$. According to a well-known SUSYQM
and SI prescription~\cite{cooper, junker, gendenshtein, dabrowska}, the latter can be recursively
determined from the ground state wavefunction (\ref{eq:gs-wf}) through the relation
\begin{equation}
  R_1^{(n)}(g,k;p) = \frac{1}{\sqrt{e_n-e_0}} b_p^+(g,k) R_1^{(n-1)}(g_1,k_1;p) \qquad n=1,
  2, \ldots.  \label{eq:recursion}
\end{equation}
Here we have specified the $(g,k)$-dependence of the operator and wavefunctions and $g_1 = g +
\beta_0$, $k_1 = k+1$, $e_0 = 0$ as consequences of (\ref{eq:ki}) and (\ref{eq:en}).\par
%
%-----------------------------------------------------------------------------------
%
In appendix 2, it is shown that the solution of equation (\ref{eq:recursion}) is given by
\begin{equation}
  R_1^{(n)}(g,k;p) = N_1^{(n)}(g,k) p^{b+\frac{1}{2}} f^{-\frac{1}{2}(a+b+1)} P^{(a,b)}_n(z)
  \label{eq:recursion-sol}
\end{equation}
where $N_1^{(n)}(g,k)$ and $P^{(a,b)}_n(z)$ are a normalization coefficient and a Jacobi polynomial,
respectively, and we have defined
\begin{equation}
  a = \frac{1}{\beta_0} \left(g - \frac{1}{2} \beta_0\right) \qquad b = k - \frac{1}{2} = j \qquad
  z = \frac{\beta_0 p^2 - 1}{1 + \beta_0 p^2} \quad (-1 < z < +1)  \label{eq:a}
\end{equation}
or, conversely,
\begin{equation}
  g = \beta_0 \left(a + \frac{1}{2}\right) \qquad k = b + \frac{1}{2} \qquad p = \left(\frac{1+z}
  {\beta_0(1-z)}\right)^{1/2} \quad (0 < p < \infty).
\end{equation}
\par
%
%----------------------------------------------------------------------------------
%
{}For $n=0$, we have shown in section 4 that the corresponding small component
radial momentum wavefunction $\trr_2^{(0)}(g,k;p)$ vanishes. For $n=1$, 2,~\ldots, and either $E_n
> 1$ or $E_n < -1$, $\trr_2^{(n)}(g,k;p)$ can be found from $R_1^{(n)}(g,k;p)$, given in
(\ref{eq:recursion-sol}), by using equation (\ref{eq:DO-eq2quater}):
\begin{equation}
  \trr_2^{(n)}(g,k;p) = \frac{\omega}{E_n+1} b_p^-(g,k) R_1^{(n)}(g,k;p).  \label{eq:small-large} 
\end{equation}
The result reads (see appendix 2):
\begin{equation}
  \trr_2^{(n)}(g,k;p) = \tnn_2^{(n)}(g,k) p^{\tb+\frac{1}{2}} f^{-\frac{1}{2}(\ta+\tb+1)}
  P^{(\ta,\tb)}_{\tn}(z)  \label{eq:small-comp}
\end{equation}
where
\begin{equation}
  \ta = a+1 \qquad \tb = b+1 \qquad \tn = n-1  \label{eq:atilde}
\end{equation}
and
\begin{equation}
  \tnn_2^{(n)}(g,k) = \frac{2\beta_0 \omega (a+b+n+1)}{E_n+1} N_1^{(n)}(g,k).
  \label{eq:norm-rel}
\end{equation}
Note that the $n=0$ case may be seen as a special case of (\ref{eq:small-comp}) by defining
$P^{(\ta,\tb)}_{-1}(z) = 0$.\par
%
%----------------------------------------------------------------------------------------
%
As stated in section 4,  $\left(R_1^{(0)}(g,k;p), \trr_2^{(0)}(g,k;p) = 0\right)$ is a (physically acceptable)
solution of equations (\ref{eq:DO-eq1quater}) and (\ref{eq:DO-eq2quater}) only for $E_0=1$. In contrast,
for any $n=1$, 2,~\ldots, $\left(R_1^{(n)}(g,k;p), \trr_2^{(n)}(g,k;p)\right)$ is a solution of
(\ref{eq:DO-eq1quater}) and (\ref{eq:DO-eq2quater}) for both $E_n > 1$ and $E_n < -1$. It remains
to check that such a solution is physically acceptable.\par
%
%----------------------------------------------------------------------------------
%
It is easy to convince oneself that the Jacobi polynomials in (\ref{eq:recursion-sol}) and
(\ref{eq:small-comp}) do not play any role in the convergence of (\ref{eq:normalization}), (\ref{eq:p2}),
which therefore does not depend on $n$. To analyse this problem it is therefore enough to consider the
smallest $n$ value, i.e., $n=0$ and $n=1$, respectively. As a consequence, the convergence is obvious for
the large component, while for the small one it directly follows from the property $\trr^{(1)}_2 \sim
R^{(0)}_1/p$ for $p \to \infty$.\par
%
%---------------------------------------------------------------------------------------
%
{}Finally, the overall normalization coefficient $N_1^{(n)}(g,k)$ can be determined from the
normalization condition of $\psi$, which amounts to equation (\ref{eq:normalization}). Use of the Jacobi
polynomial orthogonality relation~\cite{koekoek, erdelyi} then directly leads to
$N_1^{(n)}(g,k) = [(E_n+1)/(2E_n)]^{1/2} A^{(n)}(a,b)$, where $A^{(n)}(a,b)$ is given by equation
(\ref{eq:A}) in section~7.\par
%
%+++++++++++++++++++++++++++++++++++++++++++++++++
%
\subsection{Remaining cases}

In the two remaining cases corresponding to $s=\frac{1}{2}$ and large $j$ or to $s=-\frac{1}{2}$,
respectively, we can proceed similarly. In recursion relation (\ref{eq:recursion}), however, we have to
replace $g$ by $g'$ or $k$ by $k'$ in $b_p^+$ according to which case applies. In contrast, $g$ and
$k$ remain unchanged in $b_p^-$ when considering equation (\ref{eq:small-large}).\par
%
%--------------------------------------------------------------------------------------------
% 
As a result, we find that equations (\ref{eq:recursion-sol}) and (\ref{eq:small-comp}) are still valid
provided we appropriately change the definitions in (\ref{eq:a}), (\ref{eq:atilde}) and
(\ref{eq:norm-rel}), as herebelow listed:
\begin{itemize}
\item In the $s=\frac{1}{2}$ and large $j$ case, replace $g$ by $g'$ in (\ref{eq:recursion-sol})
and define $a$, $b$ $\ta$, $\tb$, $\tn$ and $\tnn_2^{(n)}(g,k)$ by
\begin{eqnarray}
  a & = & - \frac{1}{\beta_0} \left(g - \frac{1}{2} \beta_0\right) \qquad b = k - \frac{1}{2} = j \\
  \ta & = & a-1 \qquad \tb = b+1 \qquad \tn = n  \label{eq:atilde-bis} \\
  \tnn_2^{(n)}(g,k) & = & - \frac{2\beta_0 \omega (a+n)}{E_n+1} N_1^{(n)}(g',k).
       \label{eq:norm-rel-bis}  
\end{eqnarray}
\item In the $s=-\frac{1}{2}$ case, replace $k$ by $k'$ in (\ref{eq:recursion-sol}) and define $a$, $b$
$\ta$, $\tb$, $\tn$ and $\tnn_2^{(n)}(g,k)$ by
\begin{eqnarray}
  a & = & \frac{1}{\beta_0} \left(g - \frac{1}{2} \beta_0\right) \qquad b = \frac{1}{2} - k = j+1 \\
  \ta & = & a+1 \qquad \tb = b-1 \qquad \tn = n  \label{eq:atilde-ter} \\
  \tnn_2^{(n)}(g,k) & = & \frac{2\omega (b+n)}{E_n+1} N_1^{(n)}(g,k').  \label{eq:norm-rel-ter}  
\end{eqnarray}
\end{itemize}
\par
%
%----------------------------------------------------------------------------------
%
As in section~6.1, we still have to check whether the wavefunctions so obtained are physically acceptable.
Similar arguments as in 6.1 allow us to easily prove this property for the large component in both cases
and for the small component in the $s = - \frac{1}{2}$ case. For $s = \frac{1}{2}$ and large $j$, however,
we note that for $p \to \infty$, $\trr^{(0)}_2 \sim p R^{(0)}_1 \sim p^{g/\beta_0}$. As a result,
although $\trr^{(0)}_2$ is normalizable for all $g$ values satisfying condition (\ref{eq:norm-cond2}), its
contribution to $\langle p^2\rangle$ is convergent  only for
\begin{equation}
  \frac{g}{\beta_0} < - \frac{1}{2}
\end{equation}
or, in explicit form,
\begin{equation}
  2\beta \omega j - \beta' \omega > 2.  \label{eq:verylarge-j}
\end{equation}
\par
%
%--------------------------------------------------------------------------------
%
We conclude that physically acceptable solutions of equations (\ref{eq:DO-eq1quater}) and
(\ref{eq:DO-eq2quater}) exist for $n=0$, 1, 2,~\ldots\ and either positive or negative energy whenever $s
= - \frac{1}{2}$ or $s = \frac{1}{2}$ and $j$ assumes  very large values satisfying condition
(\ref{eq:verylarge-j}). In the case where $s = \frac{1}{2}$ and $j$ takes on some intermediate values in the
range
\begin{equation}
  2 - 2\beta \omega - \beta' \omega < 2\beta \omega j < 2 + \beta' \omega  \label{eq:inter-j}
\end{equation}
no physically acceptable solutions do exist.\par  
%
%=======================================================
%
\section{Final results}
\setcounter{equation}{0}

We may now collect the results obtained in sections 5 and 6 for the energy spectrum and
wavefunctions.\par
%
%----------------------------------------------------------------------------------------
%
Going back to the initial deforming parameters $\beta$, $\beta'$, we obtain for the DO complete energy
spectrum
\begin{eqnarray}
  & &E_{nsj}^2 - 1 \nonumber \\ 
  & & = 4\omega n \left[1 + \beta \omega n + \beta' \omega \left(n + j + \frac{1}{2}
       \right)\right] \qquad {\rm if\ } s = \frac{1}{2}, \ j {\rm\ small} \nonumber \\
  & & = 4\omega (n+j+1) \left[-1 + \beta \omega (n+j+1) + \beta' \omega \left(n + \frac{1}{2}
       \right)\right] \qquad {\rm if\ } s = \frac{1}{2}, \ j {\rm\ very\ large} \nonumber \\
  & & = 4\omega (n+j+1) \left[1 + \beta \omega (n+j+1) + \beta' \omega \left(n + \frac{1}{2}
       \right)\right] \qquad {\rm if\ } s = - \frac{1}{2}  \label{eq:DO-spec}
\end{eqnarray}
where $n$ runs over all values 0, 1, 2,~\ldots, except for $s=\frac{1}{2}$, $j$ small and negative
energy where $n=1$, 2,~\ldots\ only. Small and very large $j$ values are defined by conditions
(\ref{eq:small-j-bis}) and (\ref{eq:verylarge-j}), respectively. No bound states are obtained for
intermediate $j$ values satisfying condition (\ref{eq:inter-j}).\par
%
%-------------------------------------------------------------------------------
%
Equation (\ref{eq:DO-spec}) can be rewritten in a compact form 
\begin{eqnarray}
  E_{nsj}^2 - 1 & = & 4\omega \left[n + \left(1 - s - \frac{\epsilon}{2}\right) (j+1) \right] \biggl\{
       \epsilon + \beta \omega \left[n + \left(1 - s - \frac{\epsilon}{2}\right) (j+1) \right] \nonumber \\
  & & \mbox{} + \beta' \omega \left[n + \frac{1}{2} + \left(s + \frac{\epsilon}{2}\right) j \right]
       \biggr\} 
\end{eqnarray}
by introducing 
\begin{eqnarray}
  \epsilon & = & +1 \qquad {\rm if\ } s = \frac{1}{2}, \ j {\rm\ small\ or\ if\ } s = - \frac{1}{2}
       \nonumber \\
  & = & -1 \qquad {\rm if\ } s = \frac{1}{2}, \ j {\rm\ very\ large}.  \label{eq:epsilon}
\end{eqnarray}
\par
%
%------------------------------------------------------------------------------------
%
It can also be recast as
\begin{eqnarray}
  & &E_{Nsj}^2 - 1 \nonumber \\ 
  & & = 2\omega \left(N - j + \frac{1}{2}\right) \left[1 + \frac{1}{2} \beta \omega \left(N - j +
       \frac{1}{2}\right)+ \frac{1}{2}\beta' \omega \left(N + j + \frac{3}{2}\right)\right] \quad 
      {\rm if\ } s = \frac{1}{2}, \ j {\rm\ small} \nonumber \\
  & & = 2\omega \left(N + j + \frac{5}{2}\right) \left[- 1 + \frac{1}{2} \beta \omega \left(N + j +
       \frac{5}{2}\right)+ \frac{1}{2}\beta' \omega \left(N - j + \frac{3}{2}\right)\right] \nonumber \\ 
  & &  \quad {\rm if\ } s = \frac{1}{2}, \ j {\rm\ very\ large} \nonumber \\
  & & = 2\omega \left(N + j + \frac{3}{2}\right) \left[1 + \frac{1}{2} \beta \omega \left(N + j +
       \frac{3}{2}\right)+ \frac{1}{2}\beta' \omega \left(N - j + \frac{1}{2}\right)\right] \quad 
      {\rm if\ } s = - \frac{1}{2}  \label{eq:DO-specbis} 
\end{eqnarray}
or
\begin{eqnarray}
  E_{Nsj}^2 - 1 & = & 2\omega [N + 2 - s - \epsilon + (1 - 2s - \epsilon) j] \biggl\{\epsilon 
       + \frac{1}{2} \beta \omega [N + 2 - s - \epsilon + (1 - 2s - \epsilon) j]  \nonumber  \\
  & & \mbox{} + \frac{1}{2} \beta' \omega [N + 1 + s - (1 - 2s - \epsilon) j] \biggr\}
       \label{eq:DO-specter} 
\end{eqnarray}
in terms of the principal quantum number $N = 2n + l = 2n + j - s$. Here one has to take into account
that for negative energy, $s=\frac{1}{2}$ and a given $j$ value, $N$ starts at $N = j + \frac{3}{2}$
instead of $N=j-s$ as in the remaining cases.\par
%
%-----------------------------------------------------------------------------------
%
To the eigenvalues $E_{nsj}$ are associated large and small component wavefunctions defined in equations
(\ref{eq:psi-phi}), (\ref{eq:phi1}) and (\ref{eq:phi2}), where the radial momentum wavefunctions can be
written as
\begin{eqnarray}
  R_{1;s,j}^{(n)}(p) & = & \left(\frac{E_{nsj}+1}{2E_{nsj}}\right)^{1/2} A^{(n)}(a,b) p^{b+\frac{1}{2}}
         f^{-\frac{1}{2}(a+b+1)} P^{(a,b)}_n(z) \\
  R_{2;-s,j}^{(n)}(p) & = & - \epsilon\, \sigma \left(\frac{E_{nsj}-1}{2E_{nsj}}\right)^{1/2}
A^{(\tn)}(\ta,\tb)
         p^{\tb+\frac{1}{2}} f^{-\frac{1}{2}(\ta+\tb+1)} P^{(\ta,\tb)}_{\tn}(z).
\end{eqnarray}
Here 
\begin{eqnarray}
  A^{(n)}(a,b) & = & \left(\frac{2\beta_0^{b+1} (a+b+2n+1) n!\, \Gamma(a+b+n+1)}
        {\Gamma(a+n+1) \Gamma(b+n+1)}\right)^{1/2}  \label{eq:A} \\
  a & = & \frac{\epsilon}{\beta_0 \omega} \left\{1 - \beta \omega \left[\frac{1}{2} + s(2j+1)\right]
        - \frac{1}{2} \beta' \omega\right\} \qquad b = j + \frac{1}{2} - s  \\
  \ta & = & a + \epsilon \qquad \tb = b + 2s \qquad \tn = n - s - \frac{\epsilon}{2} \\
  \sigma & = & \frac{E_{nsj}}{|E_{nsj}|}
\end{eqnarray}
and $n$ runs over 0, 1, 2,~\ldots, except for $s=\frac{1}{2}$, $\epsilon = +1$ and $\sigma = -1$,
where $n=1$, 2,~\ldots.\par
%
%----------------------------------------------------------------------------------
%
{}For $\beta \omega, \beta' \omega \to 0$, it is obvious that all $j$ values fall into the ``small'' category,
as defined by equation (\ref{eq:small-j-bis}), so that equations (\ref{eq:DO-specbis}) and
(\ref{eq:DO-specter}) become
\begin{eqnarray}
  E_{Nsj}^2 - 1 & = & 2\omega \left(N - j + \frac{1}{2}\right) \qquad {\rm if\ } s = \frac{1}{2}      
       \nonumber \\
  & = & 2\omega \left(N + j + \frac{3}{2}\right) \qquad {\rm if\ } s = - \frac{1}{2} 
\end{eqnarray}
and
\begin{equation}
  E_{Nsj}^2 - 1 = 2\omega [N + 1 - s (2j+1)].
\end{equation}
Hence the deformed DO energy spectrum smoothly goes over to the nondeformed one~\cite{moshinsky89,
benitez}. In the latter, there is an infinite (resp.\ a finite) number of degenerate states with
$s=\frac{1}{2}$ (resp.\ $s = -\frac{1}{2}$), corresponding to a given $N-j$ (resp.\ $N+j$)
value~\cite{moshinsky89}. As a consequence, the conventional DO has an ${\rm so(4)} \oplus \rm so(3,1)$
dynamical symmetry Lie algebra~\cite{cq90}.\par
%
%-----------------------------------------------------------------------------------
%
{}For nonvanishing $\beta \omega$ and $\beta' \omega$ values, this degeneracy scheme is
completely spoilt so that the conventional DO symmetry Lie algebra gets broken.\par
%
%================================================
%  
\section{Conclusion}

In the present paper, we have shown that the DO problem remains exactly solvable in the momentum
representation after modifying the canonical commutation relations so that there appear isotropic
nonzero minimal uncertainties in the position coordinates. The factorization method (or, equivalently,
SUSYQM and SI techniques) has proved very convenient for deriving both the energy spectrum and the
corresponding wavefunctions of this generalized DO problem.\par
%
%-----------------------------------------------------------------------------------
%
The outcome of our study shows some resemblances to the conventional results.\par
%
%----------------------------------------------------------------------------------
%
One of them is that there is no symmetry under exchange of $s=\frac{1}{2}$ with $s=-\frac{1}{2}$.
Such a dissymetry is related to the specific substitution $\ppb \to \ppb - {\rm i} \omega \xxb \hbeta$
that has been used in constructing the DO Hamiltonian (\ref{eq:DO-eq}) and which is only one of the
two possibilities $\ppb \to \ppb \mp {\rm i} \omega \xxb \hbeta$ at our disposal. On considering
indeed equations (\ref{eq:DO-eq1quater}) and (\ref{eq:DO-eq2quater}) for $\omega$ replaced by
$-\omega$, it results from the properties $b_p^{\pm}(-\omega,s) = - b_p^{\mp}(\omega,-s)$ that
they get equivalent to the same where $s$ is changed into $-s$, and $E$, $R_1$, $\trr_2$ are
replaced by $-E$, $-\trr_2$, $R_1$, respectively. It should be stressed that in the conventional
case, the $\omega \to - \omega$ transformation has been considered in connection with
supersymmetry~\cite{cq91} or, equivalently, with the particle $\to$ antiparticle
transformation~\cite{moshinsky93a}.\par
%
%----------------------------------------------------------------------------------
% 
Another similarity, related to supersymmetry too, lies in the absence of negative-energy states with
vanishing energy $e_{0sj}=0$ (i.e., $E_{0sj}=-1$).\par
%
%----------------------------------------------------------------------------------
%
There are also some important differences between the deformed and conventional DO problems.\par
%
%---------------------------------------------------------------------------------
%
When modifying the canonical commutation relations, there appear, as expected, some additional terms
in the explicit expression for $E_{nsj}^2 - 1$, which becomes quadratic in $n$ instead of linear. As a
consequence, the degeneracy pattern observed for the conventional DO~\cite{moshinsky89, benitez}
gets completely spoilt, so that the corresponding dynamical symmetry~\cite{cq90} is broken.\par
%
%---------------------------------------------------------------------------------
%
More unexpectedly, deformation leads to a new and interesting feature in the energy spectrum: for
$s=\frac{1}{2}$, there appears a difference in behaviour between distinct $j$ values. For small ones
defined by condition (\ref{eq:small-j-bis}), the ground state has vanishing energy $e_{0sj}=0$
($E_{0sj}=1$) and SUSY is unbroken as in the conventional case. For very large $j$ values such that
condition (\ref{eq:verylarge-j}) is satisfied, the ground state acquires a nonvanishing energy
$e_{0sj}>0$ ($E_{0sj}^2>1$), as in the $s=-\frac{1}{2}$ case, so that SUSY is broken. Finally, for
intermediate $j$ values in the range (\ref{eq:inter-j}), the DO has no bound state . For deriving all these
properties, it has been essential to impose not only that the (relativistic) wavefunctions are
normalizable, but also that they lead to a finite uncertainty in $\ppb$, a supplementary condition
specific of the type of deformation considered here.\par
%
%--------------------------------------------------------------------------------
%
There remains an interesting open question in connection with the use of the deformed commutation relations
(\ref{eq:def-com}) in the DO relativistic problem. Since the conventional DO is Lorentz
covariant~\cite{moreno}, one may wonder whether such a property remains true after deformation. Although
it has been known for a long time that quantized space-time may be compatible with Lorentz
invariance~\cite{snyder}, it is clear that the formalism based on relations (\ref{eq:def-com}) is not. As
already noted before~\cite{maggiore93b}, a boost may indeed squeeze any ``minimal'' length, such as
$\Delta X_0$, as much as one likes. From a mathematical viewpoint, the question is therefore whether the
relations (\ref{eq:def-com}) can be associated with some deformed Lorentz algebra (as occurring, e.g., in the
$\kappa$-Minkowski non-commutative space-time~\cite{majid}) and, in such a case, what would be the
transformation properties of the deformed DO. From a physical viewpoint, this is related to a question of
considerable interest, which has been most debated in the recent quantum-gravity literature, namely the
possibility of Planck-scale departures from Lorentz symmetry and its experimental testing (for a recent review
see, e.g.,~\cite{amelino}).\par
%
%==============================================
% 
\section*{\boldmath Appendix 1. The operators $B^{\pm}$ in momentum representation}
\renewcommand{\theequation}{A1.\arabic{equation}}
\setcounter{section}{0}
\setcounter{equation}{0}

The purpose of this appendix is to prove equations (\ref{eq:B+}) and (\ref{eq:B-}), as well as equation
(\ref{eq:anticom}).\par
%
%----------------------------------------------------------------------------------------
%
On inserting (\ref{eq:mom-rep}) in definition (\ref{eq:B-def}) of $B^{\pm}$, the latter become
\begin{equation}
  B^{\pm} = \sigmab \cdot \pb \mp \omega \left[(1 + \beta p^2) \sigmab \cdot \frac{\partial}
  {\partial \pb} + \beta' (\sigmab \cdot \pb) \left(\pb \cdot \frac{\partial}{\partial \pb}\right) +
  \gamma \sigmab \cdot \pb\right].  \label{eq:B-momrep}
\end{equation}
\par
%
%-------------------------------------------------------------------------------------
%
To proceed further, it is convenient to apply the well-known relation
\begin{equation}
  (\sigmab \cdot \mbox{\boldmath $O_1$}) (\sigmab \cdot \mbox{\boldmath $O_2$}) = 
  \mbox{\boldmath $O_1$} \cdot \mbox{\boldmath $O_2$} + {\rm i} \sigmab \cdot
  (\mbox{\boldmath $O_1$} \times \mbox{\boldmath $O_2$}) 
\end{equation}
valid for any two vector operators $\mbox{\boldmath $O_1$}$, $\mbox{\boldmath $O_2$}$, whose
components commute with those of $\sigmab$. It leads to the following results:
\begin{eqnarray}
  (\sigmab \cdot \pb) \left(\sigmab \cdot \frac{\partial}{\partial \pb}\right) & = & \pb \cdot
       \frac{\partial}{\partial \pb} + {\rm i} \sigmab \cdot \left( \pb \times \frac{\partial}{\partial \pb}
       \right) = p \frac{\partial}{\partial p} - \sigmab \cdot \llb  \label{eq:r1}  \\
  \left(\sigmab \cdot \frac{\partial}{\partial \pb}\right) (\sigmab \cdot \pb) & = &
       \frac{\partial}{\partial \pb} \cdot \pb + {\rm i} \sigmab \cdot \left(\frac{\partial}{\partial \pb}
       \times \pb\right) = p \frac{\partial}{\partial p} + \sigmab \cdot \llb +3  \label{eq:r2}     
\end{eqnarray}
or, alternatively,
\begin{eqnarray}
  \sigma_p \left(\sigmab \cdot \frac{\partial}{\partial \pb}\right) & = & \frac{\partial}{\partial p} -
       \frac{\sigmab \cdot \llb}{p}  \\
  \left(\sigmab \cdot \frac{\partial}{\partial \pb}\right) \sigma_p & = & p \frac{\partial}{\partial p}
       \frac{1}{p} + \frac{\sigmab \cdot \llb +3}{p} = \frac{\partial}{\partial p}
       + \frac{\sigmab \cdot \llb +2}{p}      
\end{eqnarray}
where use has been made of $\sigma_p$, defined in (\ref{eq:sigmap}). Property (\ref{eq:sigmap2})
then allows one to write the operator $\sigmab \cdot \partial/\partial \pb$, appearing in equation
(\ref{eq:B-momrep}), as 
\begin{equation}
  \sigmab \cdot \frac{\partial}{\partial \pb} = \sigma_p \left(\frac{\partial}{\partial p}
  - \frac{\sigmab \cdot \llb}{p}\right) = \left(\frac{\partial}{\partial p} + \frac{\sigmab \cdot \llb
  +2}{p}\right) \sigma_p.  \label{eq:op1}
\end{equation}
\par
%
%--------------------------------------------------------------------------------------
%
{}Furthermore, the operator $(\sigmab \cdot \pb) (\pb \cdot \partial/\partial \pb)$, also appearing in
equation (\ref{eq:B-momrep}), can be expressed as
\begin{equation}
  (\sigmab \cdot \pb) \left(\pb \cdot \frac{\partial}{\partial \pb}\right) = \sigma_p p^2 
  \frac{\partial}{\partial p} = p^2 \frac{\partial}{\partial p} \sigma_p  \label{eq:op2} 
\end{equation}
as a consequence of the relation
\begin{equation}
  \frac{\partial}{\partial p} \sigma_p = \sigma_p \frac{\partial}{\partial p} 
\end{equation}
which is easily derived from equation (\ref{eq:sigmap}).\par
%
%-----------------------------------------------------------------------------------
%
On combining (\ref{eq:B-momrep}) with (\ref{eq:sigmap}), (\ref{eq:op1}) and (\ref{eq:op2}), it is then
straightforward to get equations (\ref{eq:B+}) and (\ref{eq:B-}).\par
%
%-------------------------------------------------------------------------------------
%
Let us next consider equation (\ref{eq:anticom}). From (\ref{eq:r1}) and (\ref{eq:r2}), the operator
$\sigmab \cdot \llb$ can be expressed as
\begin{equation}
  \sigmab \cdot \llb = p \frac{\partial}{\partial p} - p \sigma_p \left(\sigmab \cdot 
  \frac{\partial}{\partial \pb}\right) = \left(\sigmab \cdot \frac{\partial}{\partial \pb}\right) p
  \sigma_p - p \frac{\partial}{\partial p} - 3  
\end{equation}
from which it follows that
\begin{equation}
  \sigma_p (\sigmab \cdot \llb) = p \sigma_p \frac{\partial}{\partial p} - p \left(\sigmab \cdot 
  \frac{\partial}{\partial \pb}\right) \qquad (\sigmab \cdot \llb) \sigma_p  = \left(\sigmab \cdot
  \frac{\partial}{\partial \pb}\right) p - p \sigma_p \frac{\partial}{\partial p} - 3 \sigma_p. 
\end{equation}
These relations lead to the equation
\begin{equation}
  \{\sigma_p, \sigmab \cdot \llb\} = \left[\sigmab \cdot \frac{\partial}{\partial \pb}, p\right] -
  3 \sigma_p = - 2 \sigma_p
\end{equation}
equivalent to (\ref{eq:anticom}), which is therefore proved.\par
%
%=======================================================
%
\section*{Appendix 2. Calculation of radial momentum wavefunctions}
\renewcommand{\theequation}{A2.\arabic{equation}}
\setcounter{section}{0}
\setcounter{equation}{0}

The purpose of this appendix is to give some details on the determination of the radial momentum
wavefunctions carried out in section 6 .\par
%
%----------------------------------------------------------------------------------------
%
Let us first consider equation (\ref{eq:recursion-sol}) for $s=\frac{1}{2}$ and small $j$. On using
(\ref{eq:a}), it is clear that for $n=0$ it gives back the ground state wavefunction (\ref{eq:gs-wf}), as
it should be. Furthermore, insertion of (\ref{eq:recursion-sol}) (where $n \to n-1$, $g \to g_1$ and $k
\to k_1$) in the right-hand side of (\ref{eq:recursion}) leads to the relation
\begin{eqnarray}
  && R_1^{(n)}(g,k;p) \nonumber \\
  && = \frac{N_1^{(n-1)}(g_1,k_1)}{\sqrt{e_n}} \left[- f \frac{d}{dp} + \beta_0 \left(a + \frac{1}{2}
       \right) p - \frac{b+\frac{1}{2}}{p}\right] p^{b+\frac{3}{2}} f^{-\frac{1}{2}(a+b+3)}
       P^{(a+1,b+1)}_{n-1}(z) \nonumber \\
  && = \frac{N_1^{(n-1)}(g_1,k_1)}{\sqrt{e_n}} p^{b+\frac{1}{2}} f^{-\frac{1}{2}(a+b+1)}
       \left[- (1-z^2) \frac{d}{dz} + a - b + (a+b+2) z\right] P^{(a+1,b+1)}_{n-1}(z) \nonumber \\
  && = \frac{2n N_1^{(n-1)}(g_1,k_1)}{\sqrt{e_n}} p^{b+\frac{1}{2}} f^{-\frac{1}{2}(a+b+1)}
       P^{(a,b)}_n(z) 
\end{eqnarray}
where, in the last step, we have used equation (1.8.7) of Ref.~\cite{koekoek}. This completes the proof
that equations (\ref{eq:recursion-sol}) and (\ref{eq:a}) provide the solution of (\ref{eq:recursion}).\par
%
%-------------------------------------------------------------------------------------
%
Let us now turn ourselves to equation (\ref{eq:small-comp}). On combining (\ref{eq:small-large}) with
(\ref{eq:recursion-sol}), we obtain
\begin{eqnarray}
  && \trr_2^{(n)}(g,k;p) \nonumber \\
  && = \frac{\omega N_1^n(g,k)}{E_n+1} \left[ f \frac{d}{dp} + \beta_0 \left(a + \frac{1}{2}
       \right) p - \frac{b+\frac{1}{2}}{p}\right] p^{b+\frac{1}{2}} f^{-\frac{1}{2}(a+b+1)}
       P^{(a,b)}_n(z) \nonumber \\
  && = \frac{2\omega N_1^n(g,k)}{E_n+1} p^{b-\frac{1}{2}} f^{-\frac{1}{2}(a+b+1)}
       (1+z) \frac{d}{dz} P^{(a,b)}_n(z) \nonumber \\
  && = \frac{2\beta_0 \omega (a+b+n+1) N_1^n(g,k)}{E_n+1} p^{b+\frac{3}{2}}
       f^{-\frac{1}{2}(a+b+3)} P^{(a+1,b+1)}_{n-1}(z)
\end{eqnarray}
demonstrating the correctness of equations (\ref{eq:small-comp}) -- (\ref{eq:norm-rel}). Here, in the
last step, we have taken advantage of equation (10.8.17) of Ref.~\cite{erdelyi}.\par
%
%----------------------------------------------------------------------------------
%
In the $s=\frac{1}{2}$ and large $j$ case, equation (\ref{eq:small-large}) for the small component
radial momentum wavefunction becomes
\begin{eqnarray}
  && \trr_2^{(n)}(g,k;p) \nonumber \\
  && = \frac{\omega N_1^n(g',k)}{E_n+1} \left[ f \frac{d}{dp} + \beta_0 \left(- a + \frac{1}{2}
       \right) p - \frac{b+\frac{1}{2}}{p}\right] p^{b+\frac{1}{2}} f^{-\frac{1}{2}(a+b+1)}
       P^{(a,b)}_n(z) \nonumber \\
  && = \frac{2\beta_0 \omega N_1^n(g',k)}{E_n+1} p^{b+\frac{3}{2}} f^{-\frac{1}{2}(a+b+1)}
       \left[(1-z) \frac{d}{dz} - a\right] P^{(a,b)}_n(z) \nonumber \\
  && = - \frac{2\beta_0 \omega (a+n) N_1^n(g',k)}{E_n+1} p^{b+\frac{3}{2}}
       f^{-\frac{1}{2}(a+b+1)} P^{(a-1,b+1)}_n(z)  \label{eq:small-comp-bis}
\end{eqnarray}
on combining various properties of Jacobi polynomials~\cite{erdelyi}. Equation (\ref{eq:small-comp-bis})
agrees with equations (\ref{eq:small-comp}), (\ref{eq:atilde-bis}) and (\ref{eq:norm-rel-bis}).\par
%
%-------------------------------------------------------------------------------
%
{}Finally, in the $s=-\frac{1}{2}$ case, we get
\begin{eqnarray}
  && \trr_2^{(n)}(g,k;p) \nonumber \\
  && = \frac{\omega N_1^n(g,k')}{E_n+1} \left[ f \frac{d}{dp} + \beta_0 \left(a + \frac{1}{2}
       \right) p + \frac{b-\frac{1}{2}}{p}\right] p^{b+\frac{1}{2}} f^{-\frac{1}{2}(a+b+1)}
       P^{(a,b)}_n(z) \nonumber \\
  && = \frac{2\omega N_1^n(g,k')}{E_n+1} p^{b-\frac{1}{2}} f^{-\frac{1}{2}(a+b+1)}
       \left[(1+z) \frac{d}{dz} + b\right] P^{(a,b)}_n(z) \nonumber \\
  && = \frac{2\omega (b+n) N_1^n(g,k')}{E_n+1} p^{b-\frac{1}{2}}
       f^{-\frac{1}{2}(a+b+1)} P^{(a+1,b-1)}_n(z)
\end{eqnarray}
on combining various properties of Jacobi polynomials again~\cite{erdelyi}. Hence equations
(\ref{eq:small-comp}), (\ref{eq:atilde-ter}) and (\ref{eq:norm-rel-ter}) are valid.\par
%
%=================================================
%

\end{document}